# BAYESIAN MULTINOMIAL REGRESSION WITH CLASS-SPECIFIC PREDICTOR SELECTION[1]

By Paul Gustafson and Geneviève Lefebvre

*University of British Columbia and Université du Québec à Montréal*

Consider a multinomial regression model where the response, which indicates a unit's membership in one of several possible unordered classes, is associated with a set of predictor variables. Such models typically involve a matrix of regression coefficients, with the $(j, k)$ element of this matrix modulating the effect of the $k$th predictor on the propensity of the unit to belong to the $j$th class. Thus, a supposition that only a subset of the available predictors are associated with the response corresponds to some of the *columns* of the coefficient matrix being zero. Under the Bayesian paradigm, the subset of predictors which are associated with the response can be treated as an unknown parameter, leading to typical Bayesian model selection and model averaging procedures. As an alternative, we investigate model selection and averaging, whereby a subset of *individual elements* of the coefficient matrix are zero. That is, the subset of predictors associated with the propensity to belong to a class varies with the class. We refer to this as *class-specific predictor selection*. We argue that such a scheme can be attractive on both conceptual and computational grounds.

**1. Introduction.** Consider multinomial regression modeling where, for a single observational unit, the response variable $Y \in \{0, 1, \ldots, c\}$ indicates to which of $(c + 1)$ classes the unit belongs. Correspondingly, $X = (X_1, \ldots, X_p)'$ is a vector of $p$ predictor variables. Taking the $Y = 0$ class as a reference, the association between $Y$ and $X$ is modulated by a $c \times p$ matrix of coefficients $\beta$, with $\beta_{jk}$ describing the effect of $X_k$ on membership in the $j$th class relative to the 0th class, for $j = 1, \ldots, c$. Various link functions can be used to relate the class probabilities to the linear predictor $\beta X$.

Received July 2007; revised June 2008.

[1]Supported in part by grants from the Natural Sciences and Engineering Research Council of Canada and the Canadian Institutes for Health Research. Authors are listed alphabetically—both made substantial contributions to this work.

*Key words and phrases.* Bayesian model averaging, classification, Markov chain Monte Carlo, multinomial models.

As with other forms of regression modeling, there may be a desire or need to consider some form of $p$ predictor selection, particularly if $p$ is large. The standard approach to predictor selection is to consider eliminating some predictors entirely, which in the multinomial model context equates with setting some *columns* of $\beta$ to zero. Alternatively, though, one can consider the possibility of setting some *individual elements* of $\beta$ to zero. That is, the subset of predictors associated with the propensity of being in class $a$ may differ from the subset associated with the propensity of being in class $b$. This paper investigates this possibility and compares it with the standard approach.

To be more specific, we introduce a hierarchical prior distribution under which individual elements of $\beta$ can be zero, which we refer to as *class-specific predictor selection* (CSPS). The prior includes a hyperparameter governing the tendency for each predictor to be simultaneously irrelevant or relevant for multiple classes. A limiting value of this hyperparameter corresponds to *ordinary predictor selection* (OPS), whereby each predictor is entirely included or excluded from the model.

In fact, variants of the CSPS idea have arisen recently in the literature in different contexts. In a clustering context, both Friedman and Meulman (2004) and Hoff (2006) develop schemes whereby the subset of variables differentiating a given cluster from other clusters can vary with the cluster in question. Friedman and Meulman (2004) approach the problem by modifying existing approaches to distance-based clustering. They make the point that discovering structure whereby the subset of relevant variables differs by cluster can be important it its own right, in addition to perhaps giving a better clustering of the data overall. Hoff (2006) approaches the problem using nonparametric Bayesian technology based on Dirichlet processes, with the view that a mean-shift vector is associated to each cluster, with the position of the nonzero elements in this vector allowed to vary across clusters.

Whereas Friedman and Meulman (2004) and Hoff (2006) consider unsupervised clustering problems, the present paper deals with supervised classification problems. In an intermediate vein, Kueck, Carbonetto and de Freitas (2004) consider a semi-supervised problem, where class labels are not observed but there is some structural information about the labels in the data. These authors apply a binary regression model to relate predictors with membership in the $j$th class versus membership in any other class. They cycle through and apply this procedure to each class, that is, carry out $(c+1)$ binary regressions all told, letting the subset of relevant predictor variables be different each time. This is somewhat similar in flavor to the present approach, except that by working with multinomial regression models we obtain a real probability distribution over classes for any given value of the predictor variables. As well, the present prior specification and computational scheme bear little relation to Kueck, Carbonetto and de Freitas (2004).

Perhaps the most related work to ours is Zhou, Wang and Dougherty (2006), who do consider a form of CSPS in multinomial-probit models. However, the prior structure they use is entirely different from ours. Particularly, we use a hierarchically structured prior whereby a hyperparameter controls the *extent* to which CSPS is instantiated relative to OPS. The focus in Zhou, Wang and Dougherty (2006) is largely on computational issues surrounding scale-up to very large datasets, whereas the present emphasis is more on the interpretation and relevance of CSPS schemes. Other related work includes Sha et al. (2004) who consider (ordinary) Bayesian variable selection for multinomial-probit models. Such schemes tend to be built upon MCMC approaches to model selection and averaging in linear models [e.g., Brown, Vannucci and Fearn (1998), George and McCulloch (1993), Smith and Kohn (1996)].

In terms of modeling $X$ given $Y$ in an unsupervised context, Liu et al. (2003) and Tadesse, Sha and Vannucci (2005) pursue Bayesian schemes for (ordinary) variable selection, and Kim, Tadesse and Vannucci (2006) expand on this further using Dirichlet Process technology. Doing variable selection when modeling $X$ given $Y$ requires assumptions about how the relevant and irrelevant components of $X$ are associated, and, in fact, an independence assumption is used in the above-mentioned work. In a semi-supervised context, Gustafson, Thompson and de Freitas (2007) attempt to model $X$ given $Y$ in a way that (i) the relevant components of $X$ can vary with $Y$, and (ii) the irrelevant and relevant components of $X$ (given $Y$) are not assumed to be independent. However, this approach becomes quite cumbersome in computational terms. Issues surrounding class-specific predictor selection are much more cleanly considered in the present supervised context of modeling $Y$ given $X$.

The emphasis in what follows is that for many problems CSPS may be more scientifically plausible than OPS, as the subset of predictors which "characterize" membership in a given class will indeed vary from class to class. Thus, we argue for CSPS simply on the grounds of realistic modeling a priori. As a curious side benefit though, we find there can be computational advantages in implementing CSPS over OPS. Particularly, computational schemes for posterior sampling over a space of competing models are generally susceptible to poor performance if the models are too far apart. That is, a scheme to move from a current model to a nearby model, perhaps while keeping other parameters and latent variables fixed, may not work well due to large changes in likelihood. In moving from OPS to CSPS we induce a much larger and more nuanced space of models, that is, turning a single component of $\beta$ on or off is a much smaller change than toggling an entire column. This can facilitate the use of MCMC algorithms in computing the posterior distribution over the model space.

In Section 2 we describe the simple form of the multinomial-probit model that we use for investigating CSPS, and then describe our prior distribution which includes OPS and varying strengths of CSPS as special cases. Sections 3 through 6 contain examples of CSPS, while Section 7 gives some concluding remarks. An appendix describes implementation via MCMC techniques.

## 2. Methods.

2.1. *Multinomial-probit model.* Often multinomial models are motivated by underlying models for continuous latent variables. Particularly, the observable response $Y$ can be thought to arise from an unobservable vector of continuous variates $Z = (Z_1, \ldots, Z_c)'$, according to

$$Y = \begin{cases} y, & \text{if } Z_y = \max_j Z_j > 0, \\ 0, & \text{if } \max_j Z_j < 0. \end{cases}$$

Roughly interpreted, $Z_j$ is the propensity of the unit to belong to the $j$th class, with the biggest positive propensity "winning" the unit. If all the propensities are negative, then class zero wins. Note that the asymmetry between class zero and the other classes is actually useful in some applications, as class zero can be interpreted as those units lacking sufficient propensity to belong to any of the classes of interest.

One common specification arises from taking $Z$ to have a multivariate logistic distribution with location vector $\beta X$ (and a correlation structure arising from differencing independently distributed variates). This yields the well-known multinomial-logistic model, with

(1)
$$\frac{\Pr(Y = a|X)}{\Pr(Y = b|X)} = \exp\{(\beta_{a\bullet} - \beta_{b\bullet})X\},$$

where $\beta_{j\bullet}$ denotes the $j$th row of $\beta$ (with the convention that $\beta_{0\bullet} = 0$).

Replacing the multivariate logistic distribution with a multivariate normal distribution yields a multinomial-probit model, via

(2)
$$Z|X \sim N_c(\beta X, \Sigma).$$

As construed here, (2) has the disadvantage of not leading to a closed-form expression for the distribution of $Y|X$ akin to (1), but the advantage of a latent variable representation which is more computationally expedient. For the sake of simple exposition and computation concerning the predictor selection problem, in what follows we work with the multinomial-probit model, and, in fact, only consider the special case that the variance matrix is set to the identity matrix, that is, $\Sigma = I_c$ [Zhou, Wang and Dougherty (2006) also invoke this simplification]. This specification will be seen to facilitate computations greatly, though it also presents several limitations. First, it is a stronger assumption than is needed for identification. In fact, there is a considerable literature on identification for multinomial-probit models [see, for instance, Bunch (1991), Weeks (1997)]. Typically one can ensure formal identifiability by fixing a single diagonal element but letting $\Sigma$ be otherwise unknown. However, Keene (1992) emphasizes that "formally identified" does not always imply "practically estimable" in the multinomial-probit context. Also, Yau, Kohn and Wood (2003) argue that taking $\Sigma$ to be known is comparable in modeling flexibility to the multinomial-logistic model. Second, specifications which postulate independent latent variables are sometimes criticized on the grounds that an "independence from irrelevant alternatives" assumption is not appropriate in some applications [see Train (2003) for discussion]. Third, the specification $\Sigma = I_c$ necessarily entails some asymmetry between class zero and the other classes. A formulation under which $Z$ arises from differencing $c + 1$ independent variates (as in the multinomial-logistic model) would remove this asymmetry but would make $\Sigma$ nondiagonal. Thus, extension to nondiagonal $\Sigma$ would be a useful future development.

From the point of view of Bayesian computation, the latent variable formulation is very useful. In particular, it allows the application of MCMC to the joint posterior distribution of latent variables and parameters, which is typically more amenable than the posterior distribution of parameters alone [see, e.g., the landmark work on Bayesian multinomial models by Albert and Chib (1993)].

2.2. *Predictor selection and prior specification.* Starting with (2) as the model for the latent vector $Z$ underlying the observed $Y$, a prior distribution is specified for the matrix of coefficients $\beta$, allowing for the possibility that some of the elements are zero. Henceforth we make our notation more specific in that $X$ comprises $p$ predictors which can vary across units, whereas $\beta$ is a $c \times (p + 1)$ matrix of coefficients, with columns indexed from zero through $p$, such that the 0th column comprises intercept terms. Commensurately, let $M$ be a similarly-indexed $c \times (p + 1)$ matrix whose binary elements indicate which coefficients are nonzero. A hierarchical prior distribution of the form $\pi(\beta, M, q) = \pi(\beta | M, q)\pi(M | q)\pi(q)$ is specified, where $q$ is an unknown parameter governing the proportion of coefficients which are active, as described further below. For ease of notation the prior for $\beta$ is expressed via independent normal distributions, with zero variances for the null coefficients. That is, given $(M, q)$, the elements of $\beta$ are independently normally distributed a priori, with means and variances:

$$(3) \qquad \begin{aligned} E\{\beta_{jk} | M, q\} &= \mu_k M_{jk}, \\ \mathrm{Var}\{\beta_{jk} | M, q\} &= \begin{cases} 0, & \text{if } M_{jk} = 0, \\ M_{j+}^{-1}\tau^2, & \text{if } M_{jk} = 1. \end{cases} \end{aligned}$$

Here $M_{j+} = \sum_{k=0}^{p} M_{jk}$ denotes the number of active coefficients for the $j$th class, while $\mu = (\mu_0, \ldots, \mu_p)'$ and $\tau^2$ are user-specified hyperparameters.

As a default choice for $\mu$, the prior distribution for nonintercept terms is centered at zero, as is customary, that is, $\mu_1 = \cdots = \mu_p = 0$. Rather than taking $\mu_0 = 0$, however, we set

$$\mu_0 = \Phi^{-1}\{1 - (c+1)^{-1/c}\}. \tag{4}$$

Conceptually, setting each intercept term $\beta_{j0}$ to this value corresponds to $(Y|X_1 = 0, \ldots, X_p = 0)$ being uniformly distributed over $\{0, \ldots, c\}$. Thus, this seems an appropriate value at which to center the prior distribution of the intercept terms, particularly in problems where the predictors have been centered around zero.

The next stage of the hierarchical prior specification involves a distribution for $M$ given the unknown parameter $q$ governing the proportion of active coefficients and a hyperparameter $\rho$ which is taken as known. Here flexibility is desired, so that at one extreme all the elements of $M$ might be assumed independent a priori, which can be thought of as the strongest possible sense of CSPS, that is, the activity or inactivity of other elements in the $k$th column has no bearing on $M_{jk}$. At the other extreme, taking the column elements to be perfectly positively dependent (i.e., either all zero or all one) reverts to OPS. One simple route to achieving such flexibility in the prior distribution is described below. First though, note that the first (0th) column of $M$ is taken fixed as $M_{\bullet 0} = 1_c$, as removing intercept terms is seldom if ever sensible in multinomial regression models. Given parameter $q$, the remaining columns of $M$ are taken as independent and identically distributed. In particular, each column is assigned a mixture distribution: with probability $(1-q)$, the column elements are distributed as i.i.d. Bernoulli$\{(1 - \rho^{1/2})q\}$, and with probability $q$, they are distributed as i.i.d. Bernoulli$\{(1 - \rho^{1/2})q + \rho^{1/2}\}$. This prior specification for $M$ given $q$ (and $\rho$) is interpretable in that $\Pr(M_{jk} = 1|q) = q$ for each $(j, k)$, hence, $q$ is the "population" proportion of active coefficients. Moreover, $\mathrm{Corr}(M_{ak}, M_{bk}) = \rho$ for $a \neq b$, so that $\rho$ governs the tendency for a given predictor to be simultaneously active or inactive across the $c$ classes. In the extreme case of $\rho = 0$, all elements of $M$ are a priori independent given $q$. In the extreme case of $\rho = 1$, each column $M_{\bullet k}$ is either $0_c$ or $1_c$, that is, OPS results.

In what follows we do assign a fixed value to $\rho$, suspecting that data are not very informative about this quantity. Conversely, we do assign a prior distribution to $q$, namely, $q \sim \mathrm{Beta}(\gamma_1, \gamma_2)$. The hope is to engender some learning about $q$ from the data, that is, the data suggest what proportion of the coefficients should be active for the problem at hand.

The specification (3) gives the within-model prior variance of an active coefficient as inversely proportional to the total number of active coefficients for the class in question. This is a legitimate specification, as the prior distribution on $M$ forces each class to have at least one active coefficient (the intercept term). The rationale for this specification is as follows. Under the default choice of $\mu$, the squared distance $\|\beta_{j\bullet} - \mu\|^2$ is the sum of squared coefficients for the nonintercept coefficients for the $j$th class [plus the squared difference between the intercept term and (4)]. At least roughly then, $\|\beta_{j\bullet} - \mu\|^2$ reflects the strength of association between $Z_j$ and $X$. Now specification (3) has the particular feature that

$$E\{\|\beta_{j\bullet} - \mu\|^2 | M, q\} = \tau^2,$$

for all $M$. That is, as $M$ varies, the number of predictors involved varies while the strength of the relationship is fixed. This is seen as a desirable property, as the interpretation of $M$ is clear, and not confounded with the strength of the $Z - X$ relationship. In fact, the literature on Bayesian model averaging and selection includes discussion of how the within-model prior distribution on parameters should vary across models, and many of the suggested schemes involve the width of prior distributions for active coefficients decreasing with the number of active coefficients. See, for instance, Fernandez, Ley and Steel (2001) for discussion.

We emphasize that our prior structure is entirely different than the non-hierarchical prior employed by Zhou, Wang and Dougherty (2006). They consider only the two possibilities corresponding to $\rho = 0$ and $\rho = 1$ in our framework, and they treat the prior probability that a coefficient is active as fixed and known, rather than estimable. They also employ a different prior specification for the magnitude of active coefficients.

It should also be stressed that $M$ must be thought of somewhat differently under CSPS than under OPS, particularly in relation to the class probabilities. Under OPS, setting the $k$th column of $M$ to 0 implies the class probabilities do not depend on $X_k$. Conversely, setting $M_{jk} = 0$ under CSPS implies that the latent variable $Z_j$ does not depend on $X_k$, but necessarily $\Pr(Y = j|X)$ will still depend on $X_k$ to some extent, via the requirement that the class probabilities be normalized. It is hard then to divorce the concept of CSPS from the underlying latent variable view of multinomial regression. Thus, we interpret $M_{jk}$ as governing whether the $j$th class propensity depends on $X_k$, rather than interpreting it in terms of class probabilities.

To elaborate slightly on this issue, admittedly the CSPS interpretation is somewhat hampered by using a probit model rather than a logit model. As pointed out by a reviewer, in a multinomial-logistic model of form (1), having $M_{ak} = M_{bk} = 0$ implies that the distribution of $(Y|Y \in \{a, b\}, X)$ does not depend on $X_k$, so that some direct interpretation of the CSPS structure in terms of class probabilities is possible. However, it is straightforward to verify that this property does *not* hold in the multinomial-probit case, hence, we do tradeoff some ease of interpretation for ease of computation.

To summarize, specification of hyperparameters $\mu$, $\tau^2$, $\rho$ and $\gamma$ leads to a completely specified prior distribution. MCMC sampling from the joint posterior distribution of $(Z^{(1)}, \ldots, Z^{(n)}, \beta, M, q)$ can be implemented in a relatively straightforward manner. Ideas from Holmes and Held (2006) are useful in this regard. In particular, it is possible to marginalize and obtain an explicit expression for the prior density on $(Z, M)$, with the corresponding posterior being the truncation of this distribution to values of $Z$ consistent with the observed values of $Y$. Thus, the MCMC algorithm operates with $\pi(Z, M, q|D)$ as the target density. As a post-processing step, sampling from $\pi(\beta|Z, M, q, D)$ is used to complete the posterior sample. Further computational details are given in the Appendix.

2.3. *Inference on the matrix of coefficients.* Different approaches can be taken when estimating the coefficients $\beta$. Let $\hat{M}_{jk} = E(M_{jk}|D)$ be the posterior probability that the $(j, k)$ class-predictor pair is active, as computed via MCMC output. The first estimator of $\beta_{jk}$ we consider is simply the posterior mean, $E(\beta_{jk}|D)$. This is a weighted average of zero (with weight $1 - \hat{M}_{jk}$) and nonzero values (with weight $\hat{M}_{jk}$). On the other hand, if model selection rather than model averaging is emphasized, then one might report estimates arising from the single model $M$ having highest posterior probability. As a further alternative, following Barbieri and Berger (2004), one might report inferences arising from the median probability model $M^*(D)$, which includes exactly those terms having marginal inclusion probabilities (i.e., $\hat{M}_{jk}$) greater than 0.5. Since Barbieri and Berger (2004) present a case for good prediction arising from the median probability model, and since finding this model is an easier computational task than finding the highest posterior probability model, we consider $E(\beta|M = M^*(D), D)$ as a second estimator of $\beta$. Clearly this is a more sparse estimator than the (model-averaged) posterior mean, with potentially many components being zero.

**3. Example: synthetic data.** Data are generated with $c + 1 = 6$ classes and $p = 15$ predictors. The distribution of $(X_1, \ldots X_{15})$ is taken as multivariate normal with standardized marginals. The first block $(X_1, \ldots, X_6)$ is taken as equi-correlated with correlation coefficient 0.5. The next block is generated as $X_j \sim N(0.8X_{j-6}, 1 - 0.8^2)$, for $j = 7, \ldots, 12$, yielding across-block correlations of 0.8 for aligned elements and $0.8 \times 0.5 = 0.4$ for non-aligned elements. The remaining three elements $(X_{13}, X_{14}, X_{15})$ are taken as independent of all other elements.

In the first instance $n = 250$ observations are generated when all elements of $\beta$ are zero except for intercepts $\beta_{j0} = \Phi^{-1}\{1 - (c + 1)^{-1/c}\}$ for $j = 1, \ldots c$ (the centering value mentioned earlier), and $(\beta_{j,j}, \beta_{j,j+1}) = (0.75, 0.5)$ for $j = 1, \ldots, c$. Thus, only two coefficients are active for each class, while four predictors are relevant for exactly two classes, two predictors are relevant for a single class only, and nine predictors are completely irrelevant. In particular, predictors in the second block are completely irrelevant while being highly correlated with relevant predictors in the first block. Note that these data are generated under a scheme which matches CSPS, that is, the subset of predictors relevant for the $j$th class is small, and varies with $j$.

In the second instance again $n = 250$ observations are generated, but now the true values of $\beta$ are such that

$$|\beta_{jk}| = \begin{cases} 0.75, & \text{if } 1 \leq k \leq 3, \\ 0.5, & \text{if } 4 \leq k \leq 6, \\ 0, & \text{if } k > 6, \end{cases}$$

with different patterns of signs distinguishing the rows of $\beta$ from one another. Again the intercept terms are set equal to (4). Also, predictors in the second block are again irrelevant but highly correlated with those in the first block. This artificial situation, whereby coefficient magnitudes are constant within columns, is construed as a situation which is favorable to OPS rather than CSPS.

For both datasets posterior sampling is implemented (i) when all predictors are retained in the multinomial regression model, (ii) when CSPS is implemented with $\rho = 0$, (iii) when CSPS is implemented with $\rho = 0.7$, and (iv) with OPS (i.e., the $\rho = 1$ limit). Throughout we set $\tau^2 = 4$ as a weakly informative prior specification for probit model coefficients acting upon standardized predictors (recall that a given strength of association translates to a smaller coefficient value on the probit scale than on the more familiar and interpretable logit scale). We also set $(\gamma_1, \gamma_2) = (5, 15)$, to reflect a weak prior belief that relatively few predictors are actually active.

The ability of MCMC algorithms to mix well across the model space is invariably a concern in Bayesian model selection problems. As a simple diagnostic, two independent MCMC runs of $10^5$ iterations after $10^4$ burn-in iterations, starting from different initial models, are implemented. More precisely, the two chains are started at the empty and full models, that is, where the class-predictor pairs are either all excluded or all included. To conserve storage space, only parameter values at every tenth iteration are retained. The posterior probability of each coefficient being active, $\hat{M}_{jk} = E\{M_{jk}|D\}$, is estimated separately from the two runs, with the two sets of estimates showing good agreement (Figure 1).

To examine the MCMC mixing for $M$ in more detail, we consider *switching rates* as follows. For the thinned posterior sample of $M_{jk}$ values, the switch rate $\hat{S}_{jk}$ is simply the proportion of times that the value of $M_{jk}$ differs from its predecessor. Note that the switch rate differs from the Metropolis–Hastings acceptance rate because of the thinning, and because a change to $M_{jk}$ is proposed only once per $p$ iterations on average. The switch rate $\hat{S}_{jk}$ is plotted against $2\hat{M}_{jk}(1 - \hat{M}_{jk})$ in Figure 2, on the grounds that i.i.d. sampling of $(M|D)$ would correspond to the identity line on this plot [since the probability that two independent Bernoulli($p$) draws differ is $2p(1-p)$]. Hence, the extent to which the plotted points fall below the identity line reflects the extent to which the thinned MCMC output mixes less well than i.i.d. sampling. Particularly then, the plots reveal that the MCMC mixing is worse when $\rho = 1$ (OPS) than when $\rho < 1$ (CSPS), in line with the earlier discussion in Section 1.

For the first dataset, Figure 3 gives a graphical representation of true and estimated $\beta$ values. Generally the results from any form of predictor selection (choice of $\rho$) and either choice of estimator (posterior mean, or posterior mean conditioned on median probability model) seem preferable to the results without any predictor selection. This is particularly the case under the $\rho = 0$ prior specification, where the estimated and true forms of $\beta$ match very well. Note also that while the $\rho = 0.7$ and $\rho = 1$ fits appear similar, in the latter case the OPS scheme misses the relevance of one predictor ($X_6$) entirely.

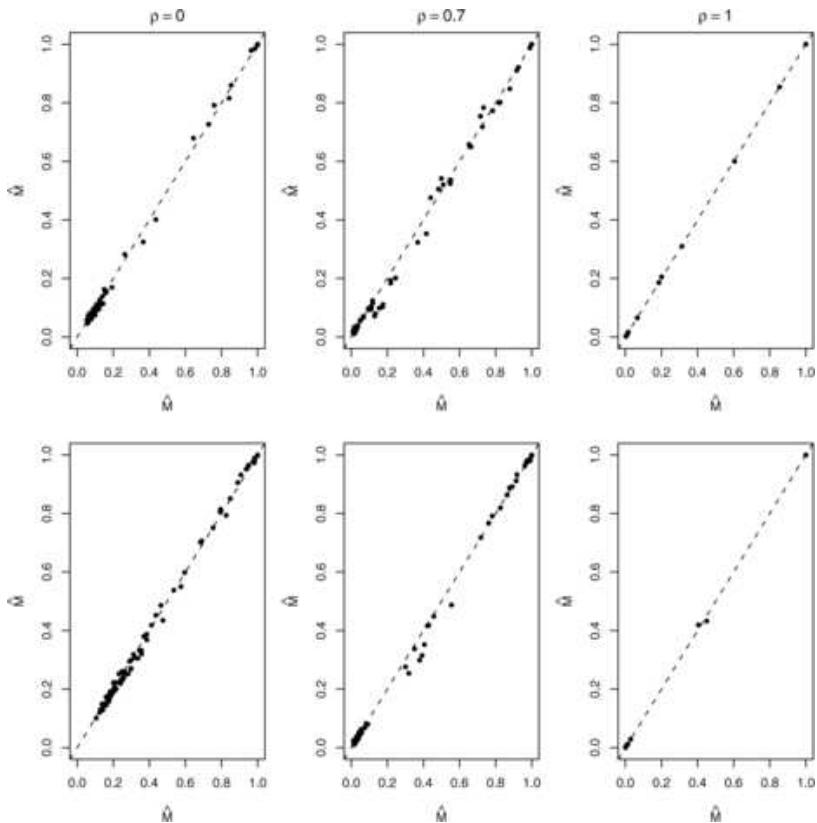

FIG. 1. *Estimates $\hat{M}$ based on two independent MCMC runs plotted against each other. The two rows of plots correspond to the two scenarios. The three columns of plots correspond to $\rho = 0$, $\rho = 0.7$, $\rho = 1$.*

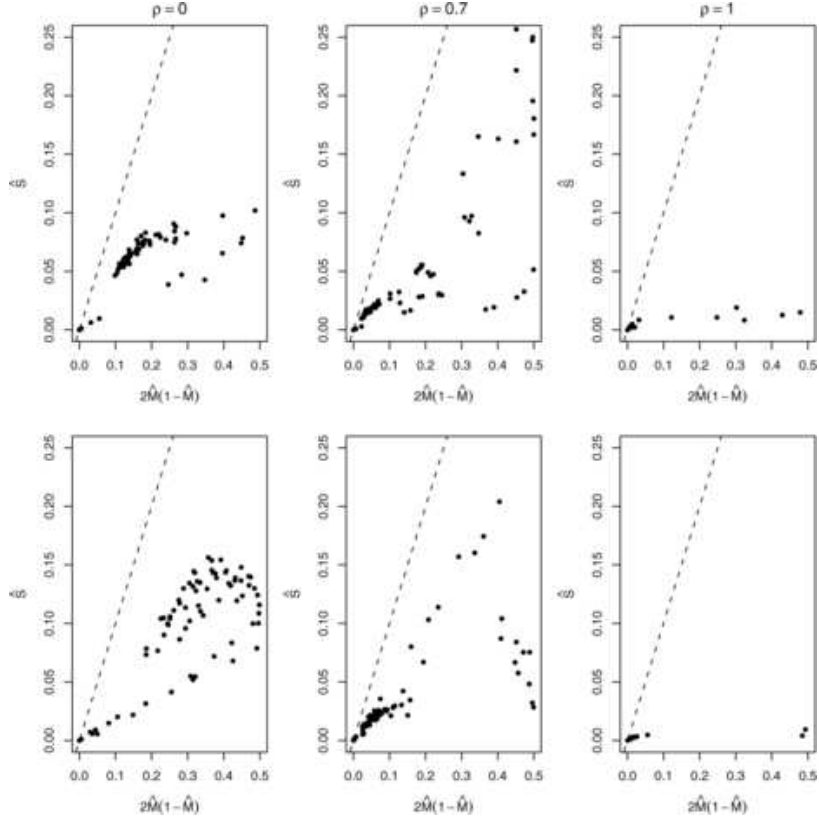

FIG. 2. *Empirical switch rates $\hat{S}$ for thinned MCMC output versus estimated switch rates under i.i.d. sampling, $2\hat{M}_{jk}(1 - \hat{M}_{jk})$. The two rows of plots correspond to the two scenarios. The three columns of plots correspond to $\rho = 0$, $\rho = 0.7$, $\rho = 1$.*

Figure 4 illustrates estimates of $\beta$ from the second dataset. Again the estimates arrived at via model selection seem preferable to those based on retaining all predictors. Even though the true $\beta$ values are compatible with OPS in this setting, the comparison between CSPS and OPS estimates seems mixed. In particular, we observe that OPS does not detect the relevance of $X_5$.

For the CSPS cases, we remark that while some relevant predictors are missed for some classes, sometimes surrogates for these predictors are detected. Although this is not desirable for interpretation purposes, it may help increase classification ability.

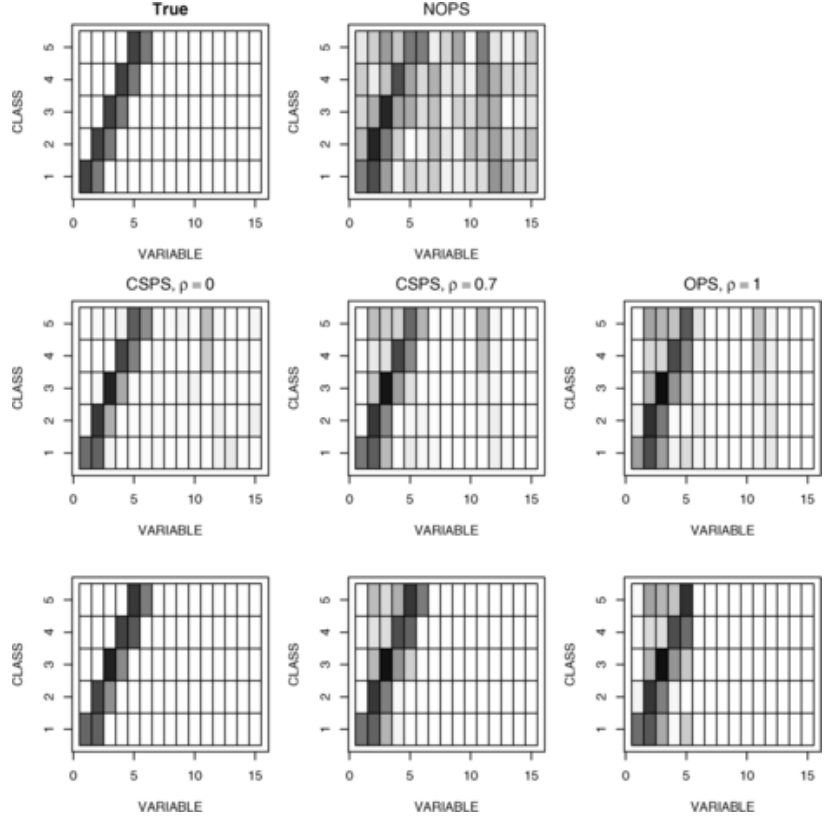

FIG. 3. *Estimation of $\beta$ for the first synthetic dataset. Each panel plots $|\hat{\beta}_{jk}|$ via a greyscale with white–black corresponding to $(0, 1.05)$. The top two panels depict true values and values estimated via including all predictors in the model. The middle row of panels corresponds to posterior mean estimation of $\beta$ under CSPS ($\rho = 0$ and $\rho = 0.7$) and OPS ($\rho = 1$). The bottom row replaces the posterior mean with the median probability model estimates.*

As a sensitivity analysis, we also implement the Bayesian analysis with a more diffuse prior on the coefficients, via $\tau^2 = 25$. Figure 5 indicates that both CSPS and OPS results are somewhat poorer in this case, in keeping with the notion that Bayesian model selection procedures can be sensitive to the choice of "within-model" prior distributions (and in the present situation the true values of coefficients are quite compatible with the narrower choice of $\tau^2 = 4$). In practice, we suggest performing sensitivity analyses to assess the degree of robustness of the results.

Finally, to examine the generality of the foregoing results (when $\tau^2 = 4$) across repeated sampling, 40 datasets are simulated under each of the two scenarios. For each dataset, the average squared error (ASE) in estimating elements of $\beta$ by their posterior means is determined, under the four modeling strategies (no predictor selection, CSPS with $\rho = 0$ and with $\rho = 0.7$, OPS with $\rho = 1$). Here the averaging of squared-error is across the $c \times (p+1) = 80$ elements of $\beta$. The distribution of ASE across repeated sampling is depicted in the boxplots of Figure 6. In Scenario 1, the advantage of some form of model selection over no selection is clear, as is the advantage of CSPS over OPS. Recall that Scenario 2 is designed as a context where OPS would likely do well, since the magnitude of elements of $\beta$ is constant within columns. Consequently, CSPS with $\rho = 0$ does not perform well relative to the other methods. However, the ASE incurred by CSPS with $\rho = 0.7$ is only slightly worse than that incurred by OPS, perhaps because MCMC sampling for CSPS tends to mix better than for OPS.

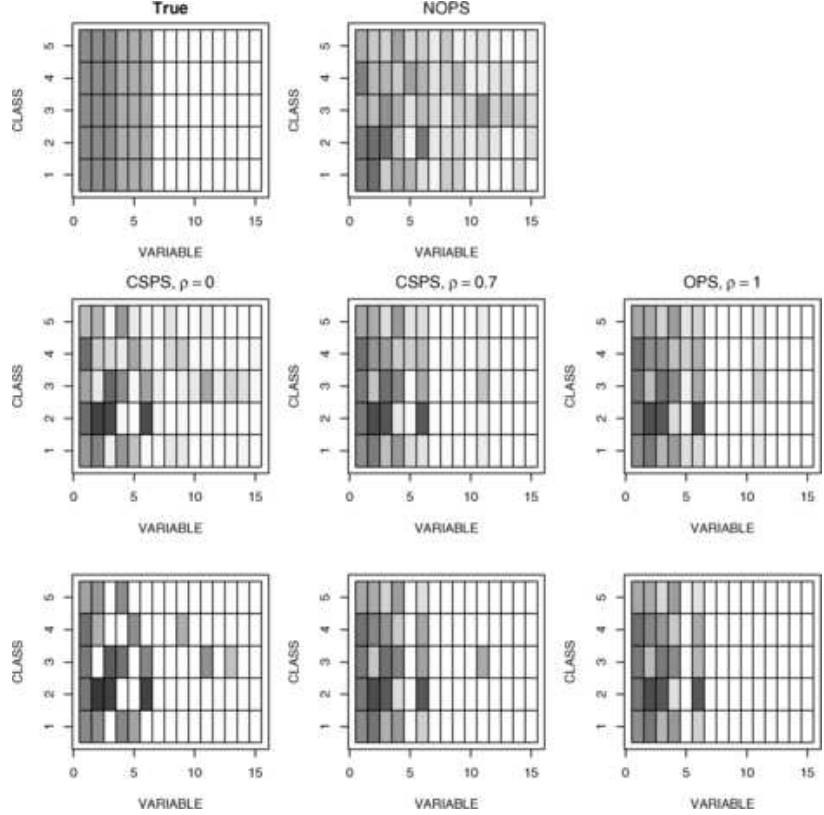

Fig. 4. *Estimation of $\beta$ for the second synthetic dataset. Each panel plots $|\hat{\beta}_{jk}|$ via a greyscale, with white–black corresponding to $(0, 1.90)$. The layout is as per Figure 3.*

**4. Example: forensic glass classification.**  We apply our method to the forensic glass dataset which has been used as an example in the classification literature [see, e.g., Ripley (1996)]. Each of $n = 214$ glass fragments belongs to one of $c + 1 = 6$ classes. The zero (reference) class is taken to be window float glass, while classes one through five are window nonfloat glass, vehicle window glass, containers, tableware and vehicle headlights respectively. The frequency distribution of fragments across classes is $(70, 76, 17, 13, 9, 29)$. The nine covariates are refractive index and percent by weight of various oxides (Na, Mg, Al, Si, K, CA, Ba, Fe). Standardized versions of these are denoted by $V = (V_1, \ldots, V_9)$.

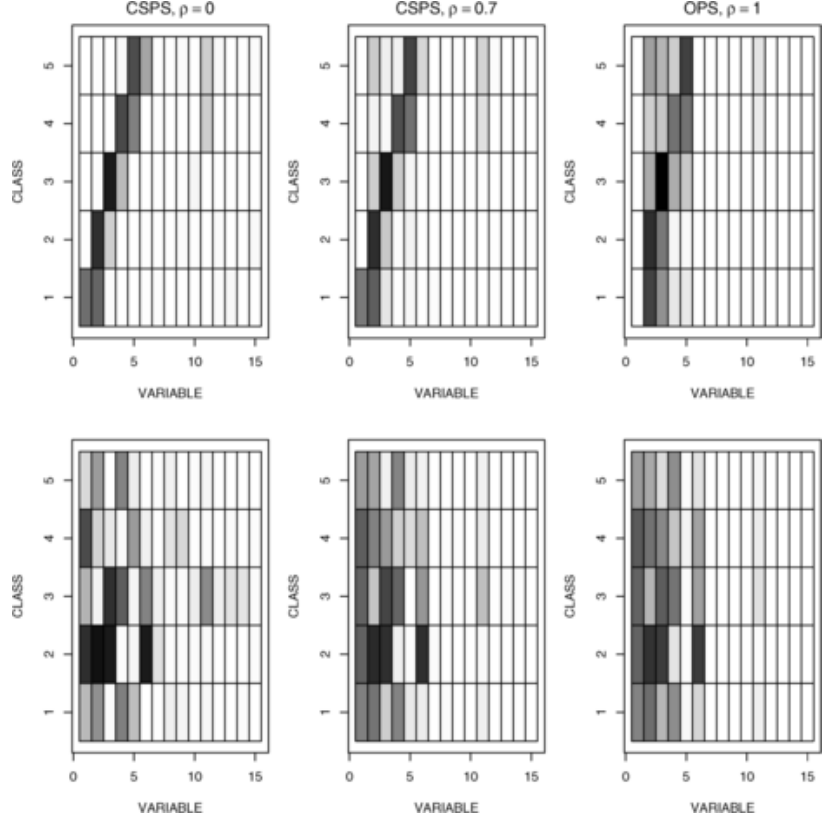

Fig. 5. *Estimation of $\beta$ under CSPS ($\rho = 0$ and $\rho = 0.7$) and OPS ($\rho = 1$) for the first and second synthetic datasets with $\tau^2 = 25$. Each panel plots $|\hat{\beta}_{jk}|$ via a greyscale, with white–black corresponding to $(0, 1.05)$ (Scenario 1); $(0, 1.90)$ (Scenario 2). The first row corresponds to estimation under the posterior mean for Scenario 1 and the second row for Scenario 2.*

To obtain a flexible model relating the covariates to the categorical response variable, we form predictors $X$ from covariates $V$ by taking radial basis functions of the form

$$(5) \qquad X_k = a_k + b_k \exp\left(\frac{-\|V - v_k^*\|^2}{2h^2}\right),$$

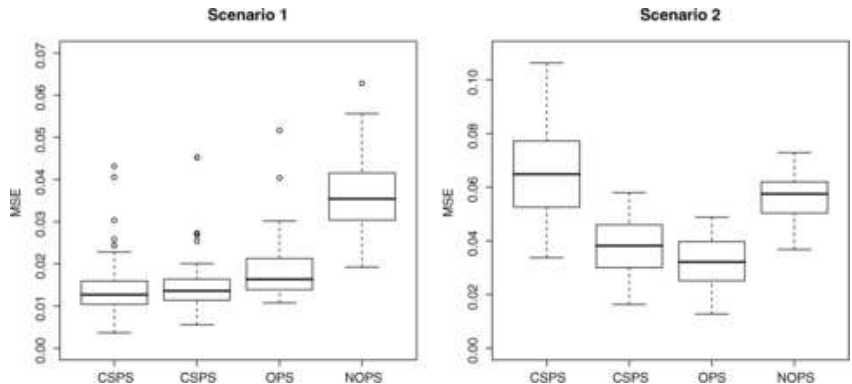

Fig. 6. *Average squared error in estimating components of $\beta$ ($\tau^2 = 4$). Each boxplot depicts the ASE for 40 simulated datasets, where ASE for a single dataset involves averaging the squared estimation error for all components of $\beta$. From left to right the methods are as follows: CSPS (with $\rho = 0$), CSPS (with $\rho = 0.7$), OPS ($\rho = 1$) and no predictor selection (NOPS), that is, all predictors included.*

for $k = 1, \ldots, p$. Radial basis functions are commonly used to obtain a flexible nonlinear model in terms of the original covariates, with the simple structure of a linear predictor for the basis functions [see, e.g., Hastie, Tibshirani and Friedman (2001), Section 6.7]. Often the knots $v_1^*, \ldots, v_p^*$ are taken to be the observed $V$ values, so that $p = n$. In the first instance, though, we set $p = 54$, and take the $v_1^*, \ldots, v_p^*$ to be $p$ randomly selected covariate values. Following Figuerido (2003), the bandwidth parameter in (5) is set at $h = 4$. The constants $a_k$ and $b_k$ are chosen to standardize $X_k$, which seems reasonable given the use of an exchangeable prior distribution across $\beta$.

In this example the hyperparameter values $\tau^2 = 25$ and $\gamma = (5, 15)$ are used, with the former specification intended to represent diffuse prior information. (Even though the empirical results in Section 3 were better when $\tau^2 = 4$, the use of a wider prior seems appropriate without specific subject-area knowledge to suggest otherwise.) For each of $\rho = 0$, $\rho = 0.7$, $\rho = 1$, two MCMC runs of $50 \times 10^4$ post burn-in iterations are implemented, with every 50th value retained. Comparison of the $\hat{M}$ estimates from the two independent MCMC runs suggests that the sampler mixes adequately over the model space (results not shown), and the remaining results are based on pooling the two runs (yielding a thinned posterior sample of size $2 \times 10^4$). The relationship between empirical switch rates $\hat{S}_{jk}$ and $\hat{M}_{jk}$ also suggests reasonable sampler performance.

For all three analyses (CSPS $\rho = 0, 0.7$ and OPS) we observe that the posterior distribution over $M$ is not favoring sparse classifiers, in that none of the class-predictor pairs have negligible posterior probability of inclusion. There is sparsity, however, in the sense that a relatively small number of components have large posterior probability. For instance, for CSPS with $\rho = 0$, the smallest $\hat{M}_{jk}$ is 0.14, yet $\hat{M}_{jk}$ exceeds 0.5 for only 9 of 270 class-predictor pairs. Larger proportions exceed this threshold for CSPS with $\rho = 0.7$ (28 out of 270 class-predictor pairs) and OPS (8 out of 54 predictors). Coefficient estimates from the three analyses appear in Figure 7.

To get some idea of the classifier performances, 10-fold cross-validation is undertaken. That is, the data are randomly split into ten equal-sized blocks, then repeatedly the CSPS and OPS models are fit to the *training* data (all but one block) to predict $Y$ from $X$ for the *validation* data (the held-out block). Note that for each split, the $p = 54$ radial basis functions are formed using covariate values randomly selected from the training data only. Point predictions are based on the mode of the predictive distribution for $(Y|X)$ (i.e., model averaging rather than selection is used). We obtain an overall misclassification rate of 29%, 29% and 30% for CSPS $\rho = 0$, CSPS $\rho = 0.7$ and OPS $\rho = 1$, respectively. These misclassification rates for CSPS are competitive with some of the results cited in the literature for these data [see, e.g., Figuredo (2003)].

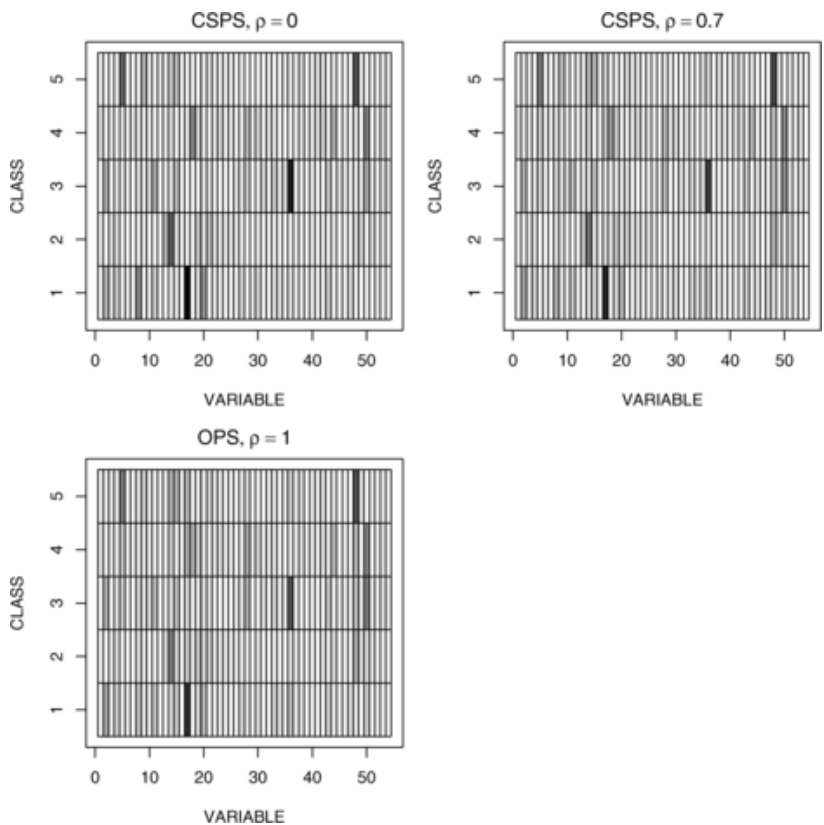

FIG. 7. *Estimation of $\beta$ for the forensic glass example. Each panel plots $|\hat{\beta}_{jk}|$ via a greyscale, with white–black corresponding to $(0, 2.40)$.*

We also investigate whether using a larger number of predictors affects the performances of the CSPS and OPS algorithms differentially. To do so, we take $p = 192$, so that when performing 10-fold cross-validation, essentially all available training set covariate values are used to form radial basis predictors. Interestingly, we now obtain an overall misclassification rate of 28%, 29%, and 42% for CSPS $\rho = 0$, CSPS $\rho = 0.7$ and OPS $\rho = 1$, respectively. While the predictive performance of CSPS is about the same as when $p = 54$, the performance of OPS clearly deteriorates. Note that the mixing of the MCMC chains for OPS is reasonable, so that OPS's decrease in performance is directly related to the model selection strategy adopted.

**5. Example: image classification.**    As a third example we consider a classification problem for which data are available from the UCI Machine Learning Repository [Asuncion and Newman (2007)]. The $n = 210$ units are $3 \times 3$ pixel regions of images of outdoor scenes, each of which is manually identified as belonging to one of seven classes (brickface, sky, foliage, cement, window, path, grass). There are 30 regions per class. The classification task is to identify the correct class using $p = 18$ continuous predictors giving attributes of the regions. These predictors involve the position of the region within the larger image, color attributes, and various other attributes (e.g., based on line detection, edge detection, etc.). A complete description of the predictors is available [Asuncion and Newman (2007)]. As in the previous examples, we standardize the predictors. Arguably image classification is a problem where it makes particular sense that the predictors useful for characterizing a given class will vary with this class. This point is also emphasized by Kueck, Carbonetto and de Freitas (2004) and Gustafson, Thompson and de Freitas (2007).

The CSPS scheme (with both $\rho = 0$ and $\rho = 0.5$), the OPS scheme and inference without predictor selection (NOPS) are applied to these data. Again, $\tau^2 = 25$ and $\gamma = (5, 15)$ are used. Visual inspection of $\beta$ estimates (plots not shown) reveals a few class-predictor pairs standing out more strongly under CSPS than under OPS or NOPS, particularly when median-posterior-model estimates are used.

To assess the impact of predictor selection on predictive performance, we generate 25 random splits of the units into training and validation samples of size 105 units each. For each training sample we fit CSPS ($\rho = 0, 0.5$), OPS and NOPS models, and generate predictive distributions for the validation sample units. Using point classifications based on the modes of the predictive distributions, Figure 8 displays the number of correct validation-set predictions for each data split and method. The results of this example are particularly interesting in that the predictive ability of the algorithm appears to be superior between the extreme cases of $\rho = 0$ and $\rho = 1$. When comparing the algorithm for these two values of $\rho$, we observe that the different data splits are in turn favoring $\rho = 0$ or $\rho = 1$. When the algorithm is run with $\rho = 0.5$, the number of correct classifications often approaches the better of the two extreme cases. Per validation sample, CSPS $\rho = 0.5$ has on average only about 0.6 more correct predictions than CSPS $\rho = 0$ and OPS (CSPS $\rho = 0$: 95.0; CSPS $\rho = 0.5$: 95.6; OPS: 94.96; NOPS: 94.64). We believe it worthy though to note that CSPS with $\rho = 0.5$ has the most stable performance across different training-validation splits.

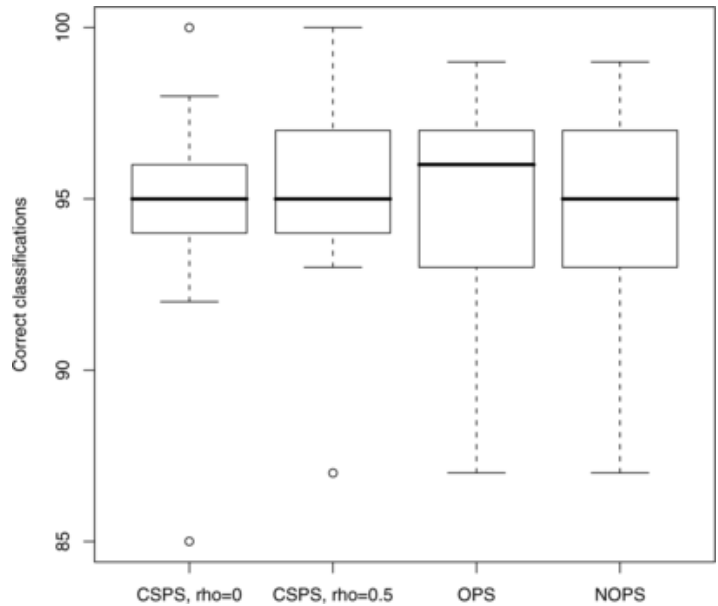

FIG. 8. *Results from 25 random splits of the image classification data. The across-split distributions of the number of correct validation set classifications (out of 105) are portrayed for CSPS ($\rho = 0, 0.5$), OPS and NOPS analyses.*

**6. Example: microarray cancer data classification.** The problem of detecting relevant genes for classification purposes from microarray data has received a lot of recent attention. The typical set-up is that among thousands of genes one tries to identify those which are relevant for predicting diagnostic categories. The fact that the number of observations is typically very small in comparison to the number of genes available often makes this task difficult. In this final example we apply our algorithm to the well-known Hereditary Breast Cancer data [Hedenfalk et al. (2001)], which can be downloaded at http://research.nhgri.nih.gov/microarray/NEJM_Supplement/. This dataset consists of 22 observations on 3226 gene expression levels. Each observation (sample) is classified into one of the following three categories: BRCA1, BRCA2 and sporadic, for a total of 7, 8 and 7 samples in each class, respectively.

As is often done with such high-dimensional problems, we initially perform a univariate screening to identify a subset of genes that are most likely to be related to the three possible outcomes. We first apply a log base-2 transformation followed by a standardization for each of the 3226 gene intensity ratios measured. We then implement a univariate CSPS analysis with $\rho = 0$ and $\tau^2 = 25$ for each of the 3226 genes, using the sporadic cancer class as reference. For each analysis, we obtain a post burn-in thinned sample for $M$ of size $2 \times 10^3$ (burn-in: 2000, thin: 10 iterations) and calculate the proportion of times the gene was selected by the algorithm for the nonreference classes BRCA1 and BRCA2. A subset of genes is selected for further analyses in the following way: we retain all genes having posterior probability of inclusion greater than 0.5 for either or both classes, for a total of 621 genes. We present in Figure 9 a histogram for the difference in gene probability of inclusion between the two nonreference classes. This histogram shows that most preselected genes are highly relevant for only one of the two classes, and therefore hints at the appropriateness of CSPS analysis over OPS.

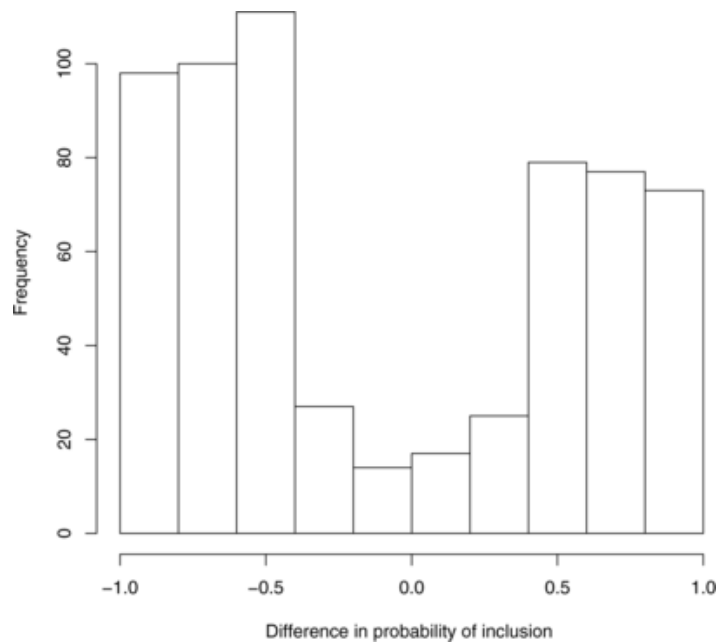

FIG. 9.  *Histogram for the difference in probability of inclusion for the preselected genes (BRCA1 vs BRCA2). Each preselected gene is such that the probability of inclusion is greater than 0.5 for at least one of the two nonreference classes.*

Many authors have analyzed this dataset. In the Bayesian literature, Yeung, Bumgarner and Raftery (2005) introduce a Bayesian model averaging algorithm where the combination of two (or more) separate logistic regressions is used as an attempt to approximate a multinomial regression model. Their algorithm finds between 13–18 relevant genes and results in 6 misclassification errors using leave-one-out cross-validation (LOOCV). Using their formulation of CSPS and OPS, Zhou, Wang and Dougherty (2006) report zero misclassification errors using a truncated version of the data and a subset of pre-screened genes.

We implement LOOCV for CSPS $\rho = 0, 0.5$ and OPS with $\tau^2 = 25$ on the subset of $p = 621$ genes. We set the prior distribution to $q \sim \text{Beta}(5, 200)$, which represents a prior belief that outcomes can be distinguished using only a very small number of genes (on average $\approx 15$ genes per class). Identifying small sets of relevant genes is generally desirable, as this is often a prerequisite for the development of inexpensive diagnostic tests [Yeung, Bumgarner and Raftery (2005)].

The fact that the algorithm allows only for one independent modification to the state of $M$ per iteration (per class) may hinder proper model exploration when the number of possible predictors is very large. To account for this feature of the algorithm, we let the MCMC run for a considerably large number of iterations. We consider 2,000,000 iterations (with a burn-in of 200,000 iterations) and retain every 200th iteration. Based on the results of the LOOCV, we obtained the posterior predictive distribution for each of the 22 samples under the model averaged estimator.

All three analyses classify correctly the same 19 of the 22 observations. In Table 1 we present the predictive probability on the true class for each observation and analysis. Note that CSPS tends to put more predictive weight on the correct class than OPS, which is desirable in the present scientific context. Moreover, there is considerable uncertainty in classifying the 5th sample under OPS, as less than 2% separates the two most likely classes. We obtain an average number of active predictors of 16 and 17 per class for CSPS $\rho = 0$ and $\rho = 0.5$, respectively, and 17 for OPS. These values are very similar to the prior expected values. In order to examine the sensitivity of the results to the choice of prior on $q$, we redo the analyses with $q \sim \text{Beta}(5, 120)$. In this case we obtain two misclassification errors for CSPS ($\rho = 0$ and $0.5$), four errors for OPS (Table 2) and an average number of active genes that similarly reflects the prior specification (25 and 26 per class for CSPS $\rho = 0$ and $\rho = 0.5$, resp., and 26 for OPS). It is not surprising that the results are highly influenced by the choice of prior on $q$, considering the very small sample size ($n = 22$) available. More experimentation with priors allowing for a larger proportion of active predictors suggests that when $p$ is large the algorithms can become impractically slow unless the prior on $q$ encourages sufficient sparsity.

*Posterior predictive probability for the correct class under LOOCV, for*
*CSPS ($\rho = 0, 0.5$) and OPS using the model averaging estimator with*
*$q \sim \text{Beta}(5, 200)$*

| Sample | CSPS $\rho = 0$ | CSPS $\rho = 0.5$ | OPS |
|--------|------|------|------|
| 1 | 0.58 | 0.56 | 0.51 |
| 2 | 0.74 | 0.73 | 0.68 |
| 3 | 0.63 | 0.58 | 0.54 |
| 4 | 0.75 | 0.75 | 0.66 |
| 5 | 0.45 | 0.42 | 0.37 |
| 6 | 0.51 | 0.50 | 0.46 |
| 7 | 0.50 | 0.52 | 0.51 |
| 8 | 0.63 | 0.60 | 0.53 |
| 9 | 0.69 | 0.82 | 0.63 |
| 10 | 0.53 | 0.52 | 0.53 |
| 11 | 0.42* | 0.42* | 0.39* |
| 12 | 0.42* | 0.39* | 0.41* |
| 13 | 0.66 | 0.65 | 0.55 |
| 14 | 0.59 | 0.58 | 0.51 |
| 15 | 0.70 | 0.69 | 0.57 |
| 16 | 0.47 | 0.46 | 0.40 |
| 17 | 0.23* | 0.23* | 0.21* |
| 18 | 0.59 | 0.56 | 0.51 |
| 19 | 0.92 | 0.92 | 0.84 |
| 20 | 0.86 | 0.86 | 0.77 |
| 21 | 0.78 | 0.73 | 0.69 |
| 22 | 0.83 | 0.86 | 0.74 |

*Misclassified observation.

**7. Discussion.** In general, with regression modeling, particularly when the number of predictors is substantial, schemes to encourage sparse solutions are called for on grounds of interpretability of the solutions, predictive generalization and computational ease. Here sparsity might refer to shrinking regression coefficients toward zero, or forcing some coefficients to be exactly zero. In the latter case, one tries to identify a subset of predictors that are judged to be irrelevant to the outcome.

In the case of multinomial response models on $(c+1)$ categories, removing a predictor from the model corresponds to setting $c$ coefficients to zero. The more nuanced strategy of CSPS that we have investigated is to set individual coefficients to zero, either without regard for the other coefficients in the same column (the $\rho = 0$ implementation of CSPS), or with some partial regard to other elements in the column ($0 < \rho < 1$). This makes conceptual sense in many applications where sparse multinomial regression solutions are sought. Moreover, at least in the examples we have considered, this

TABLE 2
*Posterior predictive probability for the correct class for CSPS*
*($\rho = 0, 0.5$) and OPS using the model averaging estimator with*
*$q \sim \text{Beta}(5, 120)$*

| Sample | CSPS $\rho = 0$ | CSPS $\rho = 0.5$ | OPS |
|--------|------|------|------|
| 1 | 0.57 | 0.56 | 0.47 |
| 2 | 0.76 | 0.74 | 0.65 |
| 3 | 0.59 | 0.63 | 0.54 |
| 4 | 0.76 | 0.75 | 0.64 |
| 5 | 0.46 | 0.44 | 0.31* |
| 6 | 0.51 | 0.51 | 0.43 |
| 7 | 0.56 | 0.50 | 0.51 |
| 8 | 0.61 | 0.66 | 0.53 |
| 9 | 0.68 | 0.76 | 0.62 |
| 10 | 0.54 | 0.54 | 0.51 |
| 11 | 0.44 | 0.44 | 0.41* |
| 12 | 0.42* | 0.43* | 0.40* |
| 13 | 0.64 | 0.68 | 0.52 |
| 14 | 0.58 | 0.59 | 0.48 |
| 15 | 0.72 | 0.71 | 0.55 |
| 16 | 0.48 | 0.48 | 0.41 |
| 17 | 0.22* | 0.23* | 0.19* |
| 18 | 0.59 | 0.58 | 0.49 |
| 19 | 0.92 | 0.92 | 0.80 |
| 20 | 0.87 | 0.84 | 0.74 |
| 21 | 0.77 | 0.76 | 0.65 |
| 22 | 0.82 | 0.83 | 0.71 |

*Misclassified observation.

conceptual sense tends to translate to estimates and predictions which are as good or better than those obtained via the standard approach of OPS.

We have also seen that to some degree computational ease is a fortunate by-product of CSPS over OPS, since it is easier to traverse a more dense space in which the constituent models are closer together. Notwithstanding this relative computational ease, however, in general, all MCMC schemes for sampling models with respect to extremely large model spaces involve a high computational burden. Indeed, one school of thought is that scaling up "true" posterior sampling over extremely large model spaces is not a realistic goal, and that MCMC algorithms are better thought of as devices to find good models [Hans, Dobra and West (2007)]. In our experience, using a C implementation [Lefebvre and Gustafson (2008)] of CSPS, computation time and mixing over $M$ is practical (from the perspective of actual posterior sampling) in problems with hundreds of predictors, but not thousands. Some modest further scale-up might be achievable through optimization of the code, but substantial further scale-up would seem to require a fresh computational approach.

## APPENDIX

In what follows we use $M[j\bullet]$ rather than $M_{j\bullet}$ to denote the $j$th row of $M$, in order to avoid nested subscripting. Also note that when a double-subscript is applied to $Z$, $Z_{ij}$ denotes the value of $Z_j$ for the $i$th unit, with $Z_{\bullet j}$ denoting the values of $Z_j$ for all units. The posterior distribution over all unknown quantities can be expressed as

$$\pi(q, M, Z, \beta | D) = \pi(q, M, Z | D)\pi(\beta | Z, M).$$

The second term $\pi(\beta | Z, M)$ involves independent multivariate normal distributions of varying dimension for the rows of $\beta$, as arises from standard Bayesian updating in normal linear models with normal priors on regression coefficients. Thus, it suffices to implement MCMC with $\pi(q, M, Z | D)$ as the target distribution. The attractive feature of this scheme is that $(q, M, Z)$ is of fixed dimension, so that straightforward MCMC updates can be applied. Specifically, we can write

$$\pi(q, M, Z | D) \propto \pi(q, M, Z) \prod_{i=1}^{n} I\{Z_{i\bullet} \in S[y_i]\}$$

$$\propto \pi(q)\pi(M | q)\pi(Z | M) \prod_{i=1}^{n} I\{Z_{i\bullet} \in S[y_i]\},$$

where $S[y]$ comprises the values of $Z = (Z_1, \dots, Z_c)$ consistent with $Y = y$. Note that $\pi(Z | M)$ takes the form

$$(6) \qquad \pi(Z | M) = \prod_{j=1}^{c} \phi_n\{Z_{\bullet j}; X_{M[j\bullet]}\mu_{M[j\bullet]}, I_n + X_{M[j\bullet]}V_{M[j\bullet]}X'_{M[j\bullet]}\},$$

where $\mu_{M[j\bullet]}$ and $V_{M[j\bullet]}$ denote the prior mean and variance for the active regression coefficients pertaining to the $j$th class. In the case that $c = 1$, Holmes and Held (2006) give an efficient algorithm to sweep through the data index and compute the mean and variance (both scalars) of $Z_i$ given $Z_{-i}$ arising from (6). This trivially extends to $c > 1$ by applying the algorithm directly (i.e., in parallel) to give the mean vector and diagonal variance matrix for $Z_{i\bullet}$ given $Z_{(-i)\bullet}$. Then for given $i$ it is a simple matter to sweep through $j$ and update $Z_{ij}$ according to the appropriate truncated normal distribution which ensures that $Z_{i\bullet} \in S[y_i]$.

To update $M$, for each $j$, a new value of $M[j\bullet]$ is proposed by toggling one component of the current value. Computation of the requisite Metropolis–Hastings acceptance probability is straightforward, again following Holmes and Held ([2006]). In particular, the relevant full conditional density takes the form

$$\pi(M[j\bullet] \mid q, M[(-j)\bullet], Z, D)$$

(7)
$$\propto |V_{M[j\bullet]}|^{1/2} |\tilde{V}_{M[j\bullet]}|^{-1/2}$$

$$\times \exp\{\tfrac{1}{2}(\mu_{M[j\bullet]} - \tilde{\mu}_{M[j\bullet]})'\tilde{V}_{M[j\bullet]}^{-1}(\mu_{M[j\bullet]} - \tilde{\mu}_{M[j\bullet]})\}$$

$$\times \pi(M|q),$$

where $\tilde{\mu}_{M[j\bullet]}$ and $\tilde{V}_{M[j\bullet]}$ are the posterior mean and variance of the active $j$th class regression coefficients. That is, $\tilde{V}_{M[j\bullet]}^{-1} = V_{M[j\bullet]}^{-1} + X'_{M[j\bullet]}X_{M[j\bullet]}$ and $\tilde{\mu}_{M[j\bullet]} = \tilde{V}_{M[j\bullet]}(X'_{M[j\bullet]}Z_{\bullet j} + V_{M[j\bullet]}^{-1}\mu_{M[j\bullet]})$.

Note that the prior density for $(M|q)$ is of complicated form, namely,

$$\pi(M|q) = \prod_{k=1}^{p}[(1-q)p_0(\rho,q)^{M_{+k}}\{1-p_0(\rho,q)\}^{c-M_{+k}}$$

$$+ qp_1(\rho,q)^{M_{+k}}\{1-p_1(\rho,q)\}^{c-M_{+k}}],$$

where $p_0(\rho,q) = (1-\rho^{1/2})q$ and $p_1(\rho,q) = (1-\rho^{1/2})q + \rho^{1/2}$. Thus, we simply use a random-walk Metropolis–Hastings algorithm to update $q$ given $(M, Z, D)$.

The above algorithm adapts quite readily to the $\rho = 1$ case of OPS. The update to $Z$ is unchanged. The update to $M$, which is now a vector, involves a product of $c$ terms of the form (7) in the acceptance probability. The update to $q$ becomes simpler, since the full conditional distribution of $q$ becomes a beta distribution.

**Acknowledgment.** We thank the editorial team for many useful suggestions.

## SUPPLEMENTARY MATERIAL

**C implementation** (DOI: [10.1214/08-AOAS188SUPP](); .zip). Code to implement CSPS and OPS appears in a supplementary material file posted at the journal website.

DEPARTMENT OF STATISTICS
UNIVERSITY OF BRITISH COLUMBIA
VANCOUVER, BRITISH COLUMBIA
CANADA V6T 1Z2
E-MAIL: gustaf@stat.ubc.ca

DEPARTMENT OF MATHEMATICS
UNIVERSITÉ DU QUÉBEC À MONTRÉAL
MONTRÉAL, QUÉBEC
CANADA H3C 3P8
E-MAIL: lefebvre.gen@uqam.ca